\documentclass[journal,twoside,web]{ieeecolor}
\usepackage{jsen}
\usepackage{cite}
\usepackage{amsmath,amssymb,amsfonts}
\newtheorem{thm}{Theorem}[section]

\newtheorem{lem}[thm]{Lemma}
\newtheorem{dfn}{Definition}[section]

\newtheorem{prf}{Proof}[section]

\newtheorem{exam}{Example}[section]

\usepackage{multirow}
\usepackage{float}
\usepackage{algorithmic}
\usepackage{algorithm}
\usepackage{array}
\usepackage{hyperref}
\usepackage[caption=false,font=normalsize,labelfont=sf,textfont=sf]{subfig}
\usepackage{textcomp}
\usepackage{stfloats}
\usepackage{url}
\usepackage{verbatim}
\usepackage{graphicx}
\usepackage{xcolor}

\usepackage{algorithmic}
\usepackage{graphicx}
\usepackage{textcomp}
\usepackage{wrapfig}
\def\BibTeX{{\rm B\kern-.05em{\sc i\kern-.025em b}\kern-.08em
    T\kern-.1667em\lower.7ex\hbox{E}\kern-.125emX}}
\definecolor{abstractbg}{rgb}{0.89804,0.94510,0.83137}
\begin{document}
\title{Compression using Discrete Multi-Level Divisor Transform for Heterogeneous Sensor Data}
\author{{Gajraj Kuldeep and Qi Zhang, \IEEEmembership{Senior Member, IEEE}}
	\thanks{This work is supported by Innovation Fund Denmark (Grant no. 8057-00059B)  and Digitalisation, Big Data and Data Analytics  center (DIGIT) Aarhus University.}
\thanks{Authors are with DIGIT, Department of Electrical and Computer Engineering, Aarhus University, Denmark.	Email:\{gkuldeep, qz\}@ece.au.dk}}

\IEEEtitleabstractindextext{%
\fcolorbox{abstractbg}{abstractbg}{%
\begin{minipage}{\textwidth}%
\begin{wrapfigure}[12]{r}{3in}%
\includegraphics[width=3in,height=1.4in]{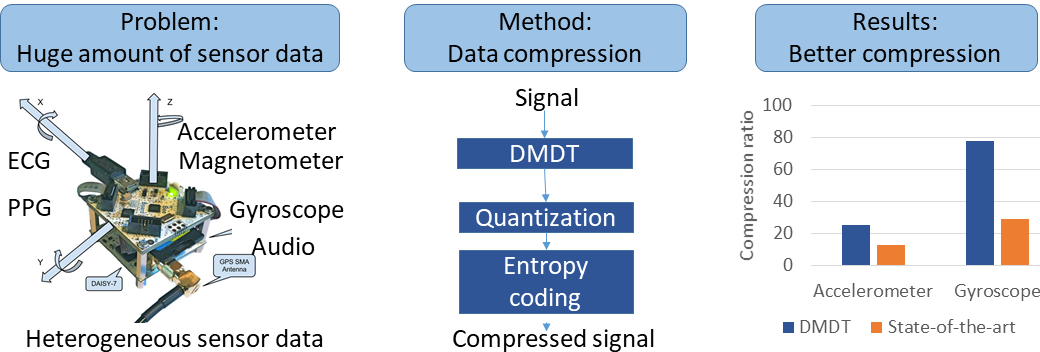}%
\end{wrapfigure}%
\begin{abstract}
	In recent years, multiple sensor-based devices and systems have been deployed in smart agriculture, industrial automation, E-Health, etc. The diversity of sensor data types and the amount of data pose critical challenges for data transmission and storage. The conventional data compression methods are tuned for a data type, e.g., OGG for audio. Due to such limitations, traditional compression algorithms may not be suitable for a system with multiple sensors. In this paper, we present a novel transform named as discrete multi-level divisor transform (DMDT). A signal compression algorithm is proposed for one-dimensional signals using the DMDT. The universality of the proposed compression algorithm is demonstrated by considering various types of signals, such as audio, electrocardiogram, accelerometer, magnetometer, photoplethysmography, and gyroscope. The proposed DMDT-based signal compression algorithm is also compared with the state-of-the-art compression algorithms.
\end{abstract}

\end{minipage}}}

\maketitle

\section{Introduction}
\IEEEPARstart{D}{uring} the last decade, the internet of things (IoT) has demonstrated an unprecedented potential to digitalize and automate diverse processes. At the same time, acquisition devices in IoT are sophisticated by incorporating many sensors in a single device, e.g., an inertial measurement unit (IMU) typically contains nine different sensors. This sophistication has enabled many IoT applications, such as predictive maintenance in Industry 4.0, autonomous vehicles, and E-Health. With the fast deployment of IoT devices with multi-sensor capabilities, a tremendous amount of data is generated. The vast volume of data poses challenges for data transmission and storage. Massive data transmission impacts the IoT device in terms of energy consumption, bandwidth, and battery lifetime. Massive data storage requires either cloud-based storage or an IoT device with huge storage capacity.

Data compression is one of the promising solutions to mitigate challenges imposed by the massive amount of data. Data compression is a well-explored area \cite{dataCom1}. It is categorized into lossless and lossy compression algorithms. Lossless compression algorithms compress data by exploiting statistical redundancy present in the data. There are many standard entropy coding methods to remove statistical redundancy from data, e.g., Huffman coding or arithmetic coding \cite{dataCom1, 5733086}. While lossy compression algorithms first employ signal transformations to exploit time or/and frequency redundancy, then entropy coding \cite{8908614}. There are many signal transforms, such as Wavelet transform and discrete cosine transform, which are used for signal transformation. Lossy data compression algorithms often provide better compression gain than lossless compression algorithms at the cost of reconstructed signal quality.    

Most data compression solutions are primarily designed to compress one type of signal from a sensor. For example, wavelet transform-based compression algorithms are customized for signals such as electrocardiogram, audio, and images \cite{8908614,9502625,NPaper}. Due to the advancement in integrated circuit technology, there are many sensors on tiny devices. In the current scenario, using a customized algorithm on such tiny devices for each signal may not be desirable due to limited memory or energy constraints. The lossless compression algorithms for different types of signals are explored in \cite{8736246,9435794}. For lossy compression, there are compression algorithms based on compressive sensing, wavelet transform, and discrete cosine transform \cite{SRISOOKSAI201237,8908614,9492306,9502626}. These algorithms exploit signal characteristics because signals are correlated in the time or frequency domain. Signals are analyzed by performing multi-resolution using transforms for example, wavelet transform \cite{8908614}, orthogonal Ramanujan periodic transform (ORPT) \cite{8469057}, and difference sequence transform (DST) \cite{9542972}. These transforms have limitations, such as wavelet transform is used for multi-resolution for radix equal to two, and bases of both ORPT and DST are not compact in the frequency domain.  

In this paper, we explore the multi-resolution for compressing different types of time-series signals. We propose a compression algorithm, which could be used for compressing different types of signals by selecting only algorithm parameters. This algorithm is designed using a novel multi-resolution transform based on discrete cosine bases named as discrete multi-level divisor transform (DMDT). The DMDT can be applied on any divisor radix, has orthogonal time bases, and has a compact representation of bases in the frequency domain. A patent application is filed for the proposed compression algorithm \cite{UDC}.   The main contributions in this paper are presented next.
\subsection{Main Contributions} Our main contributions in this paper can be summarized as
follows.
\begin{itemize}
	\item  We propose a compression algorithm based on the DMDT that can be used for various types of sensor data.
	\item We present the results for the proposed compression algorithm using signals comprising,  audio, electrocardiogram, accelerometer, magnetometer, photoplethysmography, and gyroscope. Additionally, the proposed algorithm is compared with the state-of-the-art compression algorithms using various performance metrics. The proposed algorithm achieves $55\%$ gain in compression ratio as compared to the ECG compression algorithm \cite{ECGpatent} for ECG signal. The proposed algorithm provides 2$\times$ improvement in compression ratio for accelerometer and gyroscope signal as compared to the LFZip \cite{9105816}. For audio compression the proposed algorithm provides 2-3$\times$ improvement in compression as compared to OGG \cite{Vogg}.   
\end{itemize}
\subsection{Organization and Notation}
The paper is organized as follows. Section II contains the related work. Section III presents the foundation of discrete multi-level divisor transform. Section IV explains the proposed compression algorithm.  Section V contains the performance evaluation results for the proposed algorithm for various types of time series signal.  Finally, Section VI concludes the paper. 

\textit{Notations:} In this paper, all the boldface uppercase, e.g., $\mathbf{ X}$, and all the boldface lowercase, e.g., $\mathbf{ x}$, letters represent matrices and vectors, respectively. $\mathbf{x}^T$ is transpose of $\mathbf{x}$.  Symbol $\oplus$ represents the direct sum of vector spaces. Symbol $\lfloor d \rfloor$ represents the floor of number $d$. {The $l_2$-norm of a vector $\mathbf{ x}$ is represented as $||\mathbf{ x}||_2$. Symbol $<\mathbf{ a},\mathbf{ b}>$ is the inner product of vector $\mathbf{a}$ and $\mathbf{b}$.}

\section{Related Work}
Lossless or lossy data compression methods are used to manage data explosion in the Internet of Things. The lossy compression techniques introduce a trade-off between reconstruction quality and compression ratio. Many of the lossy compression techniques use the following steps: first, transform the data in the sparse domain, then apply thresholding or quantization, and finally, entropy coding is applied. In the following, we describe the state-of-the-art lossy compression algorithms for IoT, which are often used to address data transmission and storage problems caused by multi-sensors. 


Multi-sensor data compression is required in multi-source fusion positioning system \cite{8769921}. Compressive sensing (CS) can be used to compress multi-sensor data \cite{8769921,4447198}. CS-based compression algorithms are lossy with low encoding complexity but with high reconstruction complexity \cite{9492306}.  In \cite{8769921}, data compression is performed for accelerometers, gyroscopes, magnetometers, and others. Furthermore, it is observed that for sampling rate 0.5, compression ratio (CR) is 2 with mean square error around $10^{-3}$. Our proposed algorithm has better CR at low reconstruction error than the CS-based algorithms.

Multi-sensor data compression algorithm, LFZip, for floating-point is presented in \cite{9105816}. LFZip is based on prediction of next data samples. LFZip involves prediction, quantization, and entropy coding. The prediction of  data can be performed using a filter, which can be trained beforehand using the available data. The prediction filter can use either normalized least-mean square filter or deep learning based filter. LFZip achieves better compression than SZ \cite{7516069} and critical aperture \cite{7076402}. SZ uses the best curve to predict the next point in the time-series data and critical aperture-based algorithm keeps only the subset of points depending on the error criteria. LFZip has  GPU requirements for training the predictor and realization of this predictor on resource-constrained IoT devices is challenging. In the paper, we demonstrate that the proposed algorithm achieves better compression than LFZip at similar reconstruction error.

Wavelet transforms \cite{8908614} and Ramanujan sum based transforms \cite{8469057} are used for multi-resolution and compression. There are many wavelet-based compression algorithms for different types of signals. JPEG 2000  is a wavelet transform based image compression algorithm \cite{899217}. Data compression algorithm for multi-dimensional intelligent transportation system data is proposed in \cite{7605448}. It uses wavelet transform to remove spacial and temporal correlation from the data. Wavelet transform based compression algorithms are also used for electrocardiogram signal compression \cite{NPaper,6254213}, audio and speech signal compression \cite{6254341,5972258,668558}, and accelerometer, photoplethysmography and Gyroscope signal compression \cite{8450459}. There are compression algorithms, which use discrete cosine transform instead of wavelet transform, e.g., JPEG for image compression, electrocardiogram signal compression \cite{ECGpatent}, OGG and MP3 for audio compression \cite{Vogg}, etc.

Discrete cosine transform has compact frequency basis but is not suitable for time varying signals. Wavelet transform has compact basis and is suitable for time varying signals because of the multi-resolution properties. However, Wavelet transforms are applied for radix two, i.e. each application of wavelet transform on a one-dimensional signal results in one average component and one detail component. Therefore, to obtain the next representation of a signal, wavelet transform is applied again on the average component. This process is repeated multiple times to obtain a compact representation.   

In this paper, we propose a novel discrete multi-level divisor transform (DMDT). The DMDT provides compact representation for any radix. Furthermore, we propose an compression algorithm using DMDT, that works for many different types of signals. We compare our results with the state-of-the-art compression algorithms \cite{Vogg,NPaper,ECGpatent,9105816,8769921} for various signals and demonstrate that the proposed compression algorithm achieves higher compression at better signal quality.

\section{Discrete Multi-level Divisor  Transform }
In this section, we present the formulation of the proposed discrete multi-level divisor  transform (DMDT). The following theorem lays the foundation of DMDT.

\begin{thm} \label{UDTth}
	Let $\mathbf{x}\in \mathbb{R}^N$ be a signal of length $N$. The signal length has divisor $d$ i.e., $N|d$, then the signal, $\mathbf{x}$, can be represented as:
	\begin{eqnarray} \label{UDT}
	\mathbf{s} = \mathbf{A} \mathbf{x},
	\end{eqnarray}
	where $\mathbf{A}$ is a matrix of size $N\times N$ formed by an arbitrary but fixed invertible matrix  $\mathbf{B}$ of size $d\times d$.
\end{thm}
\begin{prf}
	For a given $d$, we have  an invertible matrix  $\mathbf{B}$. Suppose this matrix has the following representation,
	\begin{equation}
	\mathbf{B} = \begin{bmatrix}
	\mathbf{ b}_1\\
	\mathbf{b}_2\\
	\vdots\\
	\mathbf{b}_d
	\end{bmatrix},
	\end{equation}
	where $\mathbf{b}_i$ is the $i^{\text{th}}$ row of the matrix $\mathbf{B}$. The matrix, $\mathbf{ A}$, is constructed using the rows of the matrix $\mathbf{B}$ as,
	\begin{equation}
	\mathbf{ A}=\begin{bmatrix}
	\mathbf{ C}_1\\
	\mathbf{ C}_2\\
	\vdots\\
	\mathbf{ C}_d
	\end{bmatrix},
	\end{equation}
	where sub-matrix $\mathbf{C}_i$ is a circulant matrix, which is constructed using the $i^{\text{th}}$ row of the matrix $\mathbf{B}$, i.e., row vector $\mathbf{b}_i$.
	The $i^{\text{th}}$ circulant sub-matrix $\mathbf{C}_i$ is constructed first, by extending the $i^{\text{th}}$ row vector $\mathbf{b}_i$ to length $N$ by appending $N-d$ zeros. After that the extended vector is shifted by  $md$, where $m =0, 1,2 \dots, \frac{N}{d}-1$ to get $\frac{N}{d}$ vectors.  The matrix constructed using $\mathbf{b}_i$ is given as,
	
	\begin{equation}
	\mathbf{ C}_i=\begin{bmatrix}
	\mathbf{b}_i & \mathbf{ 0}_{1\times d} & \mathbf{ 0}_{1\times d} &  \dots & \mathbf{ 0}_{1\times d} \\
	\mathbf{ 0}_{1\times d} & \mathbf{b}_i & \mathbf{ 0}_{1\times d} &  \dots & \mathbf{ 0}_{1\times d}\\
	\vdots & \vdots \\
	\mathbf{ 0}_{1\times d} & \mathbf{ 0}_{1\times d}  & \mathbf{ 0}_{1\times d} &  \dots & \mathbf{b}_i\\
	\end{bmatrix}_{ \frac{N}{d} \times N}.
	\end{equation}
	This shifting operation results in $\frac{N}{d}$ orthogonal vectors for each row vector. Therefore, the total number of row vectors of the matrix $\mathbf{ A}$ is $N$ for a given matrix  $\mathbf{B}$.  	
	Hence the matrix, $\mathbf{ A}$ is invertible from construction. This proves that the signal $\mathbf{x}$ can be represented by $\mathbf{ s}$ using Eq. \ref{UDT}. \\
\end{prf}
The theorem \ref {UDTth} is generic and depends on the construction of the matrix $\mathbf{B}$. For example, for $d=2$ and the matrix $\mathbf{B}$ is constructed using one low-pass filter and one high-pass filter, and it constructs the family of  Wavelet transforms.
%
\begin{exam}
	For $d=2$, Haar Wavelet matrix, $\mathbf{A}$, is constructed using the following matrix,
	\begin{eqnarray*}
		\mathbf{B} = \left[\begin{array}{rr}1 &1\\
			1 & -1\end{array}\right].
	\end{eqnarray*}
	The application of the matrix $\mathbf{A}$ on a signal $\mathbf{ x}$ will result in an average component and a detail  component.
\end{exam}

Wavelet theory is studied extensively for $d=2$ and applied on various signal processing applications, including compression, multiresolution, etc. \cite{PPV,WT}. A novel divisor based multi-band Wavelet transform has been designed using Ramanujan Sums in \cite{8469057} and difference sequences \cite{9542972}. We present the following example from \cite{8469057} for $d=3$.
\begin{exam}
	For $d=3$, the  matrix, $\mathbf{A}$, is constructed using the following matrix,
	\begin{eqnarray*}
		\mathbf{B} = \left[\begin{array}{rrr}1 & 1 & 1\\
			2 & -1 & -1\\
			0 & 1 & -1\end{array}\right].
	\end{eqnarray*}
	The application of the matrix $\mathbf{A}$ on a signal $\mathbf{ x}$ will result into an average component and two detail  components.
\end{exam}

By following similar nomenclature,  from Eq. \ref{UDT} a signal $\mathbf{x}$ is decomposed into $d$ components. Next, we describe the proposed DMDT by applying Eq. \ref{UDT} multiple times on the first component in the following lemma.   

\begin{lem} \label{decomposition}
	For a given signal $\mathbf{ x}$ of length $N$ with  $N_1=N$, there is $N_{k+1}=N_{k}|d_{k}$. For each $d_k$ there is an invertible orthogonal matrix $\mathbf{B}_k$, which is used to construct matrix $\mathbf{ A}_k$ using theorem \ref{UDTth}. For the  divisor, $d_1$, the decomposed signal $\mathbf{s}_1=\mathbf{ A}_1\mathbf{ x}$ can be represented as,
	\begin{equation} \label{equ5}
	\mathbf{ s}_1=\mathbf{v^1}\oplus\mathbf{w^1_1}\oplus\dots\oplus \mathbf{w^1_{d_1-1}},
	\end{equation}
	where $\mathbf{v^1}$ is the average component,  $\mathbf{w^1_i}$s are the detail components after applying signal decomposition using divisor $d_1$, and $\oplus$ is direct sum. Now the component  $\mathbf{v^1}$ can be further decomposed into $\mathbf{s}_2=\mathbf{ A}_2\mathbf{ v}^1$ using divisor $d_2$ and it can be represented as,
	\begin{equation}\label{equ6}
	\mathbf{s }_2=\mathbf{v^2}\oplus\mathbf{w^2_1}\oplus\dots\oplus \mathbf{w^2_{d_2-1}},
	\end{equation}
	where $\mathbf{v^2}$ is  the average component and $\mathbf{w^2_i}$s are the detail components after applying signal decomposition using divisor $d_2$. Similarly for the $k^\text{th}$  decomposition using divisor $d_k$, $\mathbf{s}_{k}=\mathbf{ A}_{k}\mathbf{ v}^{k-1}$ is given as,
	\begin{equation}\label{equ7}
	\mathbf{s_{k}}=\mathbf{v^k}\oplus\mathbf{w^k_1}\oplus\dots\oplus \mathbf{w^k_{d_k-1}}.
	\end{equation} 
	
	The final decomposed signal $\mathbf{z}$ is constructed using the $k^{\text{th}}$ average component and all the detail components from first level to the $k^{\text{th}}$ level. The final decomposed signal, $\mathbf{ z}$, can be represented as,
	\begin{multline}
	\mathbf{ z}=\mathbf{v^k}\oplus\mathbf{w^k_1}\oplus\dots\oplus \mathbf{w^k_{d_k-1}}\oplus \mathbf{w^{k-1}_1}\oplus\dots \\\oplus \mathbf{w^{k-1}_{d_{k-1}-1}}\oplus \dots \oplus \mathbf{w^1_1}\oplus\dots\oplus \mathbf{w^1_{d_1-1}}.
	\end{multline}
\end{lem}
\begin{prf}
	Using the fact that matrix $\mathbf{ B}_k$ is orthogonal for  $d_k$. From theorem \ref{UDTth}, we know that sub-matrices constructed using rows of the  $\mathbf{ B}_k$ are orthogonal to each other. The  $k^\text{th}$ matrix $\mathbf{ B}_k$ and constructed matrix $\mathbf{ A}_k$ can be represented as,    
	\begin{equation}
	\mathbf{B}_k = \begin{bmatrix}
	\mathbf{ b}^k_1\\
	\mathbf{b}^k_2\\
	\vdots\\
	\mathbf{b}^k_d
	\end{bmatrix}  \text{ and }
	\mathbf{ A}_k=\begin{bmatrix}
	\mathbf{ C}^k_1\\
	\mathbf{ C}^k_2\\
	\vdots\\
	\mathbf{ C}^k_d
	\end{bmatrix},
	\end{equation}
	where, $\mathbf{ C}^k_j$ is the $j^\text{th}$ circulant sub-matrix constructed using the   $j^\text{th}$ row, i.e., $\mathbf{b}^k_j$ of the matrix $\mathbf{B}_k$. From construction, we have $<\mathbf{ C}^k_l,\mathbf{ C}^k_n>=0$ for all $k$ and $l\ne n$. This shows that the each sub-matrix constructs an orthogonal sub-space. This proves that these sub-spaces can be represented using direct sum as in Eq. \ref{equ5}, \ref{equ6}, and \ref{equ7}, depending on the divisor. Finally, the final decomposed vector is a concatenation of the decomposed vectors from each level.  
\end{prf}

\begin{figure}[]
	\centering
	\subfloat[\label{cc1}]{\includegraphics[width=\linewidth]{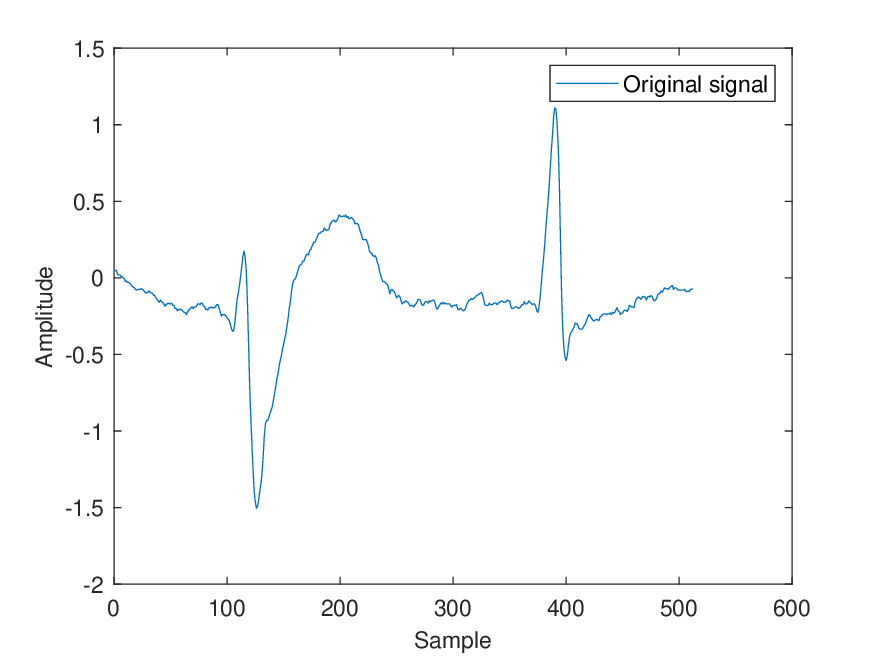}}\\
	\subfloat[\label{cc2}]{\includegraphics[width=\linewidth]{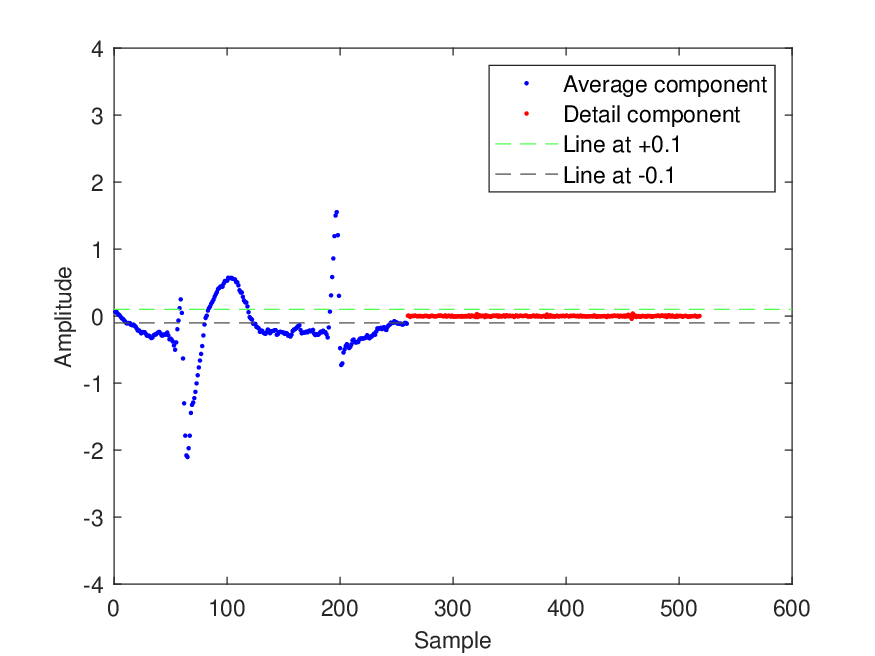}}\\
	\subfloat[\label{cc3}]{\includegraphics[width=\linewidth]{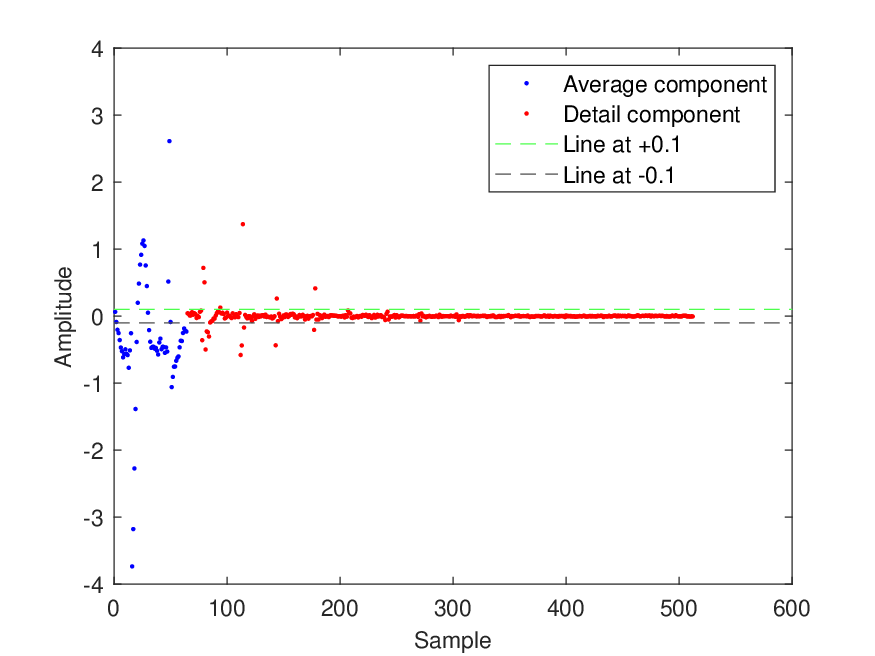}}
	\caption{ (a) ECG signal of length 512. (b) Transformed signal after applying Wavelet transform  on ECG signal with symmlet filter of length 8. (c) Transformed signal after applying DMDT on  ECG signal with divisor 8. } 
	\label{ecgFig}
	
\end{figure}
%

The DMDT for one-dimensional signals depends on the construction of the matrix, $\mathbf{B}$, for a divisor $d$. There are many choices for $d$. For example, when $d=2$, Wavelet bases are orthogonal in the time domain and are compact in the frequency domain such as  Daubechies wavelet, Morlet wavelet, Cohen–Daubechies–Feauveau (CDF) 9/7, etc.  For $d\ge 2$ Ramanujan sum based  wavelet transform are defined and its bases are orthogonal in time domain. However, they are not compact in the frequency domain. 

In this paper, we use discrete cosine bases for constructing the matrix $\mathbf{B}$, because discrete cosine bases are orthogonal in time domain and compact in frequency domain even for $d>2$. We use the following definition to construct the rows of the matrix, $\mathbf{ B}$ for the divisor $d$. Note that DMDT is not limited to discrete cosine bases, and one could use different construction matrix, e.g., modified discrete cosine bases \cite{545218} or selected analysis and synthesis filters \cite{1164954}.

\begin{dfn}
	For any  integer, $d\ge2$, the entries, i.e., $b_{nm}$ of the matrix $\mathbf{ B}$, are defined as, 
	\begin{equation}
	b_{nm}=cos\left (\frac{\pi n(2m+1)}{2d}\right), 
	\end{equation}
	where $0\le n \le d-1$ and $0\le m \le d-1$. 
\end{dfn}

A few examples of the matrix $\mathbf{B}$ for $d=2$ and $d=3$ are given as,

\begin{equation*}
\mathbf{ B}=\begin{bmatrix}
1 & 1\\
0.7071 & -0.7071
\end{bmatrix}\text{   } 
\mathbf{ B}=\begin{bmatrix}
1 & 1 & 1\\
0.8660 & 0 &-0.8660\\
0.5 & -1 & 0.5
\end{bmatrix}.
\end{equation*}

We demonstrate why DMDT is superior as compared to the wavelet transforms by presenting more compact representation in the transform domain, by taking electrocardiogram (ECG) as an example. Fig. \ref{ecgFig} shows the comparison of the  wavelet transform and DMDT for compact representation of an ECG signal. Original ECG signal of length 512 is shown in Fig. \ref{cc1}. Transformed signal using WT is shown in Fig. \ref{cc2}. This compactness is observed by applying WT using Symmelet filter of length 8. DMDT transformed signal is shown in Fig. \ref{cc3}, which is obtained using only one level of decomposition using divisor 8. In this example, DMDT is constructed using normalized discrete cosine bases.  To show compactness of the transformed signals for wavelet transform and DMDT, two lines are indicated at $+0.1$ and $-0.1$. It can be observed that the DMDT provides more compact signal representation, i.e., the fewer number of average coefficients in the transform domain. Additionally, there are more transformed coefficient, which are close to zero in DMDT than WT.  

{We also compare the computational complexity of the DMDT and Wavelet transform. For DMDT the filter length and radix are the same, i.e., $d$. However, for Wavelet transform it is not the case, and generally radix and filter length are different. We take a simple case where the DMDT and Wavelet transform have the same filter length. For filter length is $d$, both the DMDT and  Wavelet transform require $(d-1)N$ additions and $dN$ multiplications. For this case the length of the average component is $\frac{N}{d}$ and the length of the combined detail components is $\frac{(d-1)N}{d}$  for the DMDT. Similarly for the Wavelet transform the length of the average component is $\frac{N}{2}$ and the length of the detail component is also $\frac{N}{2}$. It can be observed that $\frac{(d-1)N}{d}>\frac{N}{2}$ for $d> 2$. This shows that when $d>2$ the DMDT takes the same number of additions and multiplications as Wavelet transform but provides more compact representation, i.e., transformed signal values are close to zero, which can be observed from Fig. \ref{cc3}. Now we study the case where wavelet transform and DMDT are applied once more on the corresponding average components. First we take the wavelet transform, applying wavelet transform for the second time gives the average component of length $\frac{N}{4}$ and  the detail component  of length $\frac{N}{4}$ with  $(d-1)\frac{N}{2}$ additions and $d\frac{N}{2}$ multiplications. Similarly, DMDT gives the average component of length $\frac{N}{d^2}$ and  the detail component  of length $\frac{(d-1)N}{d^2}$ with  $(d-1)\frac{N}{d}$ additions and $N$ multiplications. It can observed that the DMDT takes less number of additions and multiplication for $d> 2$. We also give the following example to show that wavelet is applied three times to achieve similar signal compactness as compared to DMDT. For example to achieve similar compactness as in Fig. \ref{cc3}, Wavelet transform is applied three times and will require 7*(512+256+128) additions and 8*(512+256+128) multiplications whereas the DMDT will only require 7*512 additions and 8*512 multiplications for similar compactness as shown in Fig. \ref{cc3}. Therefore the DMDT is more suitable for real-time resource-constrained applications.} 

Till now,  DMDT is shown for one-dimensional signal. Now we present the case for two-dimensional signal. The first level decomposition of a two-dimensional signal can be performed by constructing two matrices. For example, let $\mathbf{{X}}$ be a two-dimensional signal of size $N\times L$ with $N|d_{N_1}$ and $L|d_{L_1}$. The transformation of  a signal $\mathbf{X}$ is given as, $\mathbf{ S=RXH}^T,$ where the matrix $\mathbf{R}$ of size $N\times N$ and $\mathbf{ H}$ of size $L\times L$ are constructed using divisor $d_{N_1}$ and $d_{L_1}$ using theorem \ref{UDTth}. Multi-level decomposition for two-dimensional signals can be obtained similar to the one-dimensional case as in lemma \ref{decomposition}. We consider the case when divisors for each dimension are same for multi-level decomposition and present the following lemma.

\begin{lem} \label{decomposition1}
	For a two-dimensional signal $\mathbf{ X}$ of size $N\times L$ with $N_1=N$ and $L_1=L$, there are $N_{k+1}=N_{k}|d_{k}$ and $L_{k+1}=L_{k}|d_{k}$. For each $d_k$ there is an invertible orthogonal matrix $\mathbf{B}_k$, which is used to construct matrix $\mathbf{ R}_k$ and $\mathbf{ H}_k$ using theorem \ref{UDTth}. For  divisor $d_{1}$, the decomposed signal $\mathbf{ S}_1=\mathbf{ R}_1 \mathbf{ X} \mathbf{ H}^T_1$ can be represented as,
	\begin{equation}
	\mathbf{ S}_1=\mathbf{V^1}\oplus\mathbf{W^1_1}\oplus\dots\oplus \mathbf{W^1_{d^2_1-1}},
	\end{equation}
	where $\mathbf{V^1}$ is the average component and $\mathbf{W^1_i}$s are the detail components after applying decomposition using divisor $d_1$. Now the component  $\mathbf{V^1}$ can be decomposed into $\mathbf{ S}_2=\mathbf{ R}_2 \mathbf{ V}^1 \mathbf{ H}^T_2$ using divisor $d_2$ and it is represented as,
	\begin{equation}
	\mathbf{ S}_2=\mathbf{V^2}\oplus\mathbf{W^2_1}\oplus\dots\oplus \mathbf{W^2_{d^2_2-1}},
	\end{equation}
	where $\mathbf{V^2}$ is the average component and $\mathbf{W^2_i}$s are the detail components after applying decomposition using divisor $d_2$. Similarly for the $k^\text{th}$ decomposition using divisor $d_k$, $ \mathbf{ S}_k=\mathbf{ R}_k \mathbf{ V}^{k-1} \mathbf{ H}^T_k$ is given as,
	\begin{equation}
	\mathbf{ S}_{k}=\mathbf{V^k}\oplus\mathbf{W^k_1}\oplus\dots\oplus \mathbf{W^k_{d^2_k-1}}.
	\end{equation} 
	
	The final decomposed signal $\mathbf{ Z}$ can be represented as,
	\begin{multline}
	\mathbf{ Z}=\mathbf{V^k}\oplus\mathbf{W^k_1}\oplus\dots\oplus \mathbf{W^k_{d^2_k-1}}\oplus \mathbf{W^{k-1}_1}\oplus\dots\\\oplus \mathbf{W^{k-1}_{d^2_{k-1}-1}}\oplus \dots \oplus \mathbf{W^0_1}\oplus\dots\oplus \mathbf{W^0_{d^2_1-1}}.
	\end{multline}
\end{lem}
\begin{prf}
	Proof is similar to the proof of lemma \ref{decomposition}.
\end{prf}
In the next section, we present the proposed compression algorithm for one-dimensional signals.

\begin{figure*}
	\centering
	\subfloat[\label{comAlgo1}]{\includegraphics[width=.8\linewidth]{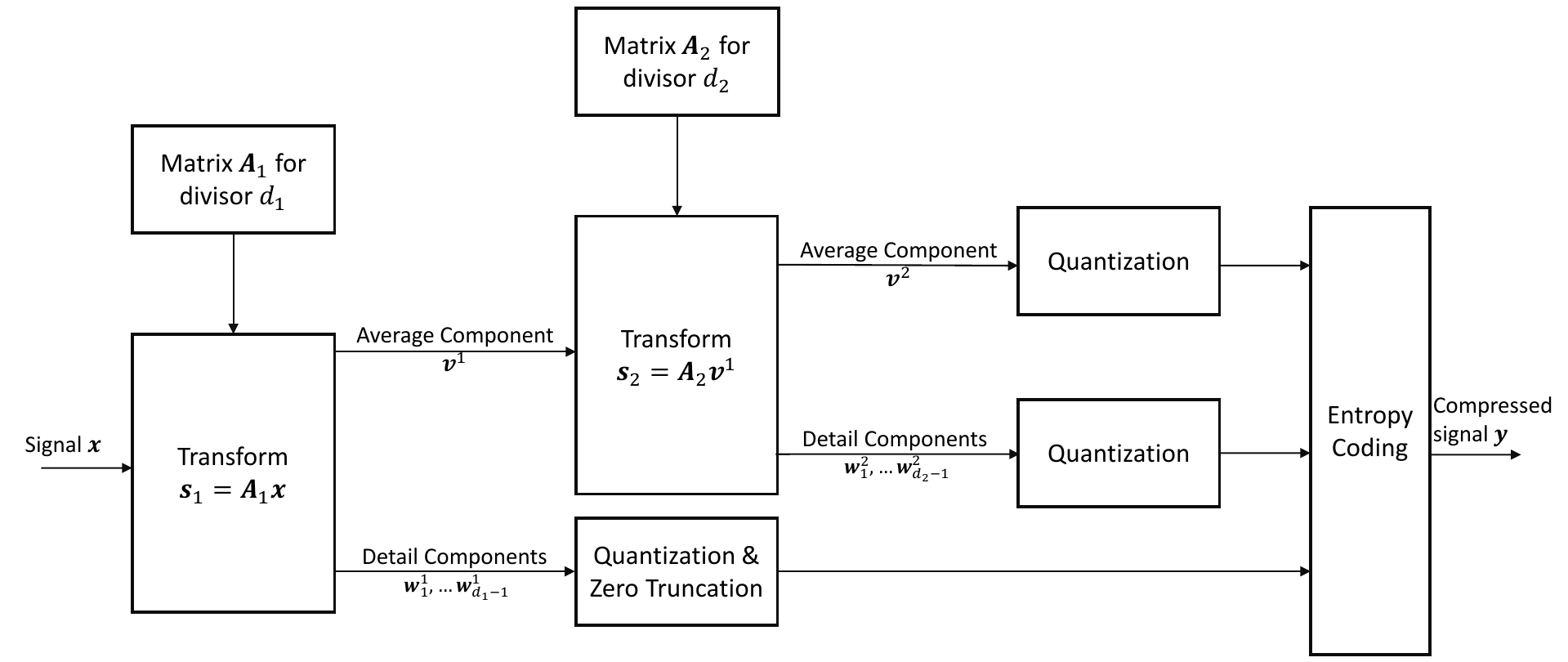}}\\
	\subfloat[\label{decomAlgo1}]{\includegraphics[width=.8\linewidth]{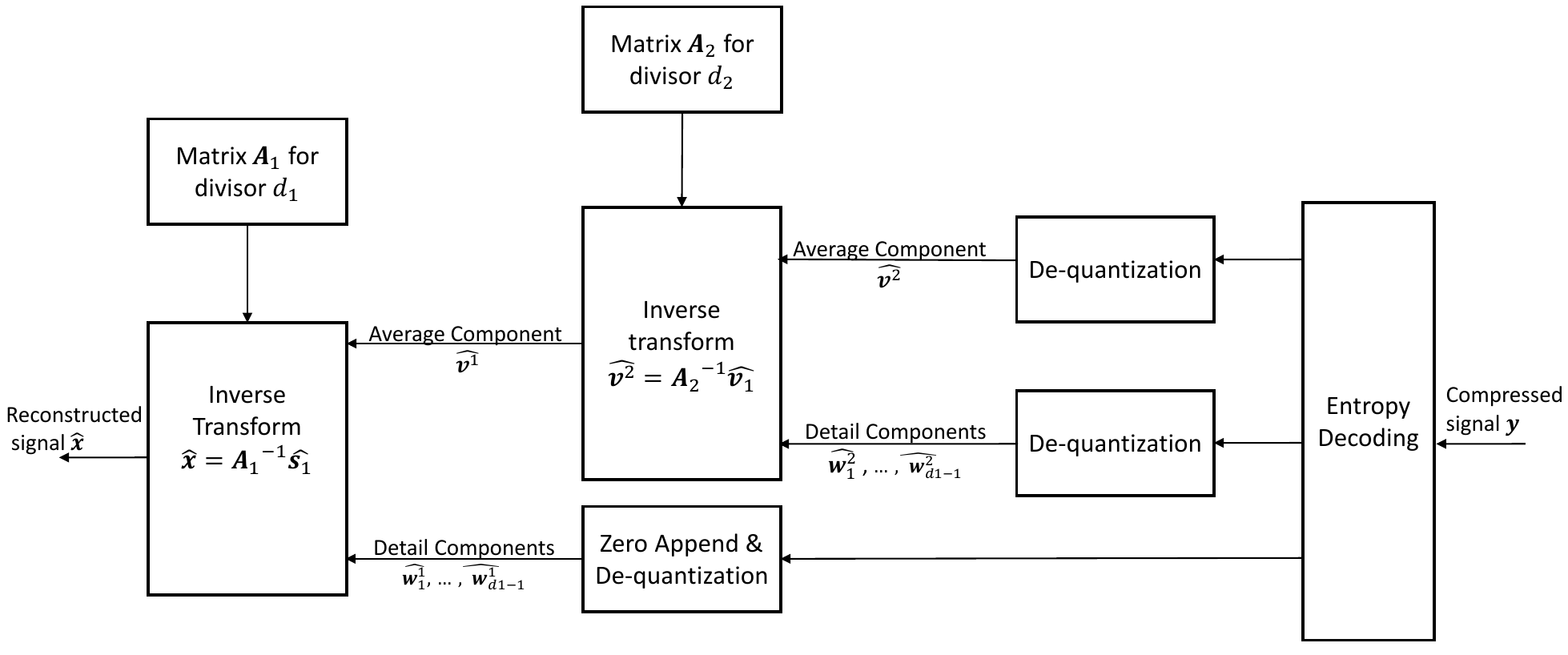}}
	\caption{(a) Block diagram of the one-dimensional compression algorithm using DMDT. (b) Block diagram of the one-dimensional decompression algorithm.}
	\label{comDecom}
\end{figure*}

\section{Compression Algorithm}
In this section, we present a compression algorithm named as 1D-DMDT. This compression algorithm is based on DMDT as proved in lemma \ref{decomposition}. From the lemma there are two parameters to apply DMDT on a signal. First, the number of levels, i.e., the number of times DMDT is applied. Second, the divisors for each level. In this paper, we take two levels of DMDT and two divisors for DMDT matrix construction. The block diagrams of the proposed compression and decompression algorithm are shown in Fig. \ref{comDecom}.  In this algorithm the divisors are selected based on signal length, i.e., first divisor satisfies $N|d_1$ and second divisor satisfies $(N|d_1)|d_2$. From Fig. \ref{comAlgo1}, it can be observed that after applying DMDT two times, we get one average component of length $\frac{N}{(d_1d_2)}$ and $d_1+d_2-2$ detail components of length $N-\frac{N}{(d_1d_2)}$.

{ After applying DMDT two times, all the components are passed through a quantization step. The quantization step is denoted as $Q$. In this quantization step we use a normalization constant $\theta$, whose value is positive. The quantization parameter, $\theta$, trade offs the quality of the reconstructed signal and compression ratio. The lower value of $\theta$ provides good reconstructed signal quality with lower compression ratio, whereas the higher value of $\theta$ provides high compression with lower reconstructed signal quality. Quantization step is the only step in the proposed compression algorithm that introduces the artefacts in the reconstructed signal.} The  quantization step for the second step average component is performed as,
\begin{equation}
Q(\mathbf{v}^2)=\lfloor\mathbf{v}^2\frac{1}{\theta\sqrt{d_1 d_2}}+0.5\rfloor,
\end{equation}
where $\mathbf{ v}^2$ is the average component after applying DMDT for the second time and $\frac{1}{\sqrt{d_1 d_2}}$ is a normalization factor. 

The quantization step for the second level detail components is given as,
\begin{equation}
Q(\mathbf{w}^2_i)=\lfloor\mathbf{w}^2_i\frac{\sqrt{2}}{\theta\sqrt{d_1 d_2}}+0.5\rfloor,
\end{equation}
where $\mathbf{ w}^2_i$ is the $i^\text{th}$ detail component after applying DMDT for the second time and $\frac{\sqrt{2}}{\sqrt{d_1 d_2}}$ is a normalization factor.

For the detail components of the first level signal decomposition, the quantization step can be performed using,
\begin{equation}
Q(\mathbf{w}^1_i)=\lfloor\mathbf{w}^1_i\frac{\sqrt{2}}{\theta\sqrt{d_1}}+0.5\rfloor,
\end{equation}
where $\mathbf{ w}^1_i$ is the $i^\text{th}$ detail component of the first level signal decomposition and $\frac{\sqrt{2}}{\sqrt{d_1}}$ is a normalization factor. Finally the second level quantized average component, the second level quantized detail components, and the first level quantized detail components are serialized and passed to an entropy encoder to remove statistical redundancy. Note that if the signal has positive samples then the mean value is first subtracted from the second level average component, after that it is quantized and the mean value is added to the compressed data that will be used in decompression. {Entropy coding, which is employed to eliminate statistical redundancy from the data, is utilized in both lossless and lossy compression. There are many types of entropy encoders, such as the Huffman encoder, Arithmetic encoder, and Asymmetric numeral systems. In the proposed compression algorithm, the final step is entropy coding, with arithmetic coding being employed \cite{515510}. In the proposed algorithm, one could opt for other entropy coding methods depending on the use case. For instance, if the speed of compression and decompression is crucial, then Huffman coding can be chosen over arithmetic coding. A detailed explanation of entropy coding can be found here \cite{dataCom1, 5733086}.}  The final entropy coded data is represented using $\mathbf{y}$.

Decompression procedure is shown in Fig. \ref{decomAlgo1}. The first step in decompression is entropy decoding and de-serialization, i.e., converting back to the average and detail components. In de-quantization of the second level average component, if the mean value is present then after de-quantization the mean value is added back, and  after that de-quantization is performed detail components using the knowledge of divisors and normalization constant. Finally, inverse transform is applied depending on the number of levels. 

Three parameters in the proposed compression algorithms, i.e., the number of signal decompositions, the divisor for each signal decomposition, and the normalization constant, determine compression ratio and signal reconstruction quality. They can be tuned and configured depending on the application. In the next section, we show the empirical effect of these parameters on signal quality and compression ratio on various types of time-series signals. 

%

\section{Performance Evaluation}

\subsection{Datasets and Performance Metrics }
For evaluating the proposed algorithm, we use several different time-series datasets. Datasets used in this paper comprise electrocardiogram (ECG) data, photoplethysmograph (PPG) data, audio data, smartwatch sensor data, power consumption data, and IMU data. The ECG dataset is taken from  MIT-BIH ECG database \cite{dbECG1,932724}. This dataset contains 48 ECG records, and each record is sampled at $360$ Hz and represented in 11-bit resolution. Wearable dataset \cite{10.1145/3242969.3242985} with the following modalities is used: blood volume pulse, electrodermal activity, respiration, and body temperature. 	Photoplethysmography (PPG) dataset using wireless sensor for activity
monitoring is taken from \cite{BIAGETTI2020105044}. Data recorded in this dataset are from 7 subjects and includes 105 PPG signals (15 for each subject) and the corresponding accelerometer signals measured with a sampling frequency of 400 Hz with 16-bit resolution \cite{BIAGETTI2020105044}. We use magnetometer and microphone recording from \cite{CMU1,CMU2}. We also apply the proposed algorithms on the datasets from the UCI Machine Learning Repository \cite{Dua:2019}. From this repository, heterogeneity human activity recognition \cite{10.1145/2809695.2809718} is used for simulation. This dataset uses 36 smartphones and smartwatches consisting of 13 different device models. Data in the dataset are sampled using sampling frequency from 25 Hz to 200 Hz.    

The proposed work is applied on a wide range of signals. Therefore, we use the most applicable and widely used metrics for measuring the quality and compression performance. In the following, we define these metrics.

Reconstruction performance of ECG signals is often measured using percentage root mean square difference (PRD) between the original signal and reconstructed signal, which is given as, 
\begin{equation}
\text{PRD}=(\sqrt{\frac{||\mathbf{ x}-\mathbf{ \hat{x}}||_2^2}{||\mathbf{ x}||_2^2}})100,
\end{equation}
where $\mathbf{ x}$ is the original signal and $\mathbf{ \hat{x}}$ is the reconstructed signal. For measuring the compression, we use compression ratio (CR), which is given as,
\begin{equation}
\text{CR}=\frac{\text{Number of bits in original signal }}{\text{Number of bits in compressed signal}}.
\end{equation}
We also use quality score (QS) for showing the trade-off between the reconstruction performance and compression ratio for ECG signals. The QS is defined as,
\begin{equation}
\text{QS}=\frac{\text{CR}}{\text{PRD}}.
\end{equation}

The reconstruction performance is also measured using signal to noise ratio (SNR) for audio, accelerometer, magnetometer, etc., which is defined as,
\begin{equation}
\text{SNR}=10 log(\frac{||\mathbf{ x}||^2_2}{||\mathbf{ x-\hat{x}}||^2_2}).
\end{equation}

\subsection{Electrocardiogram and Photoplethysmography}
In this subsection, we apply the proposed algorithm on ECG signals and compare with wavelet-based compression methods. ECG signals are used from dataset \cite{dbECG1,932724}. In the simulation, both small signal length and large signal length scenarios are considered. Small signal length can be realized in resource-constrained IoT device for real-time applications to reduce transmission cost, latency, and bandwidth usage. Large signal length scenario is useful for long term storage. The simulation results are shown in Table \ref{ECGT1} and \ref{ECGT2}.  In Table \ref{ECGT1} simulation is performed by applying DMDT two times with $d_1=32$ and $d_2=16$. To demonstrate the effect of normalization constant on compression ratio, various values of $\theta$ are considered. In Table \ref{ECGT2} the normalization constant is fixed in order to achieve PRD of 0.47. Simulation results presented in Table \ref{ECGT2} show the effect of signal length on the CR and QS. Note that, compression algorithm in \cite{ECGpatent} is designed for small signal lengths, whereas compression algorithm in \cite{NPaper} is designed for large signal length. From Tables \ref{ECGT1} and \ref{ECGT2}, it can be observed that the 1D-DMDT compression algorithm not only performs better for small signal lengths but also for large signal lengths. From Tables \ref{ECGT1}, it can be observed that the 1D-DMDT algorithm achieves $55\%$ gain in CR as compared to the ECG compression algorithm from \cite{ECGpatent} for ECG signal at similar QS values of around 29. 
\begin{table}[t!]
	\centering
	\caption{Performance comparison of the 1D-DMDT and the state-of-the-art schemes for ECG signal. For 1D-DMDT signal length is 512 and first level divisor is 32 and second level divisor is 16. }\label{ECGT1}\vspace{0.2cm}
	\resizebox{\columnwidth}{!}{
		\begin{tabular}{c|c|c|c|c|c|c|}
			\cline{2-7}
			& \multicolumn{4}{c|}{1D-DMDT} & Ref. \cite{ECGpatent} & Ref. \cite{NPaper}\\ \hline
			\multicolumn{1}{|c|}{}    & $\theta$=5  &  $\theta$=10 &  $\theta$=15 &  $\theta$=20  &        &        \\ \hline
			\multicolumn{1}{|c|}{CR}  & 4.60  & 7.06  & 8.61  & 10.05 & 5.25   & 1.62   \\ \hline
			\multicolumn{1}{|c|}{PRD} & 0.13  & 0.22  & 0.29  & 0.35 & 0.19   & 0.47   \\ \hline
			\multicolumn{1}{|c|}{QS}  & 34.17 & 32.78 & 29.75 & 28.37  & 29.93  & 3.44   \\ \hline
	\end{tabular}}
\end{table}

\begin{figure}[htbp]
	\centering
	{\includegraphics[width=.8\linewidth]{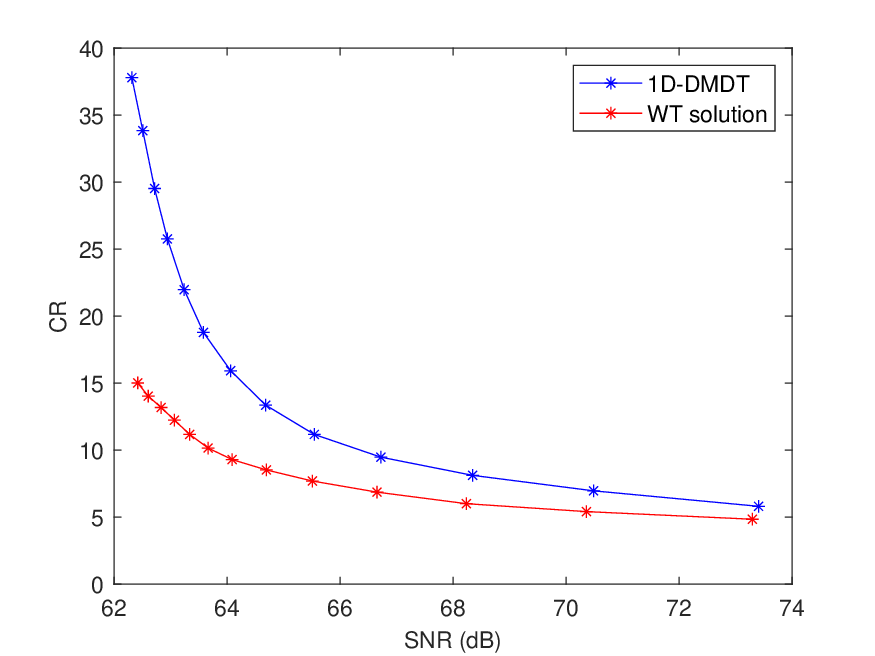}}

	\caption{Performance of 1D-DMDT using PPG signals for $N=512$ and various values of thresholds.  For 1D-DMDT first level divisor is 32 and second level divisor is 16. Dataset \cite{BIAGETTI2020105044}}
	\label{ppg1}
\end{figure}

\begin{table}[t!]
	\centering
	\caption{Performance comparison of the 1D-DMDT and \cite{NPaper} for ECG signal with various signal lengths at PRD of 0.47. For 1D-DMDT threshold is 30, and first level divisor is 32 and second level divisor is 16. } \label{ECGT2}\vspace{0.2cm}
	\resizebox{\columnwidth}{!}{
		
		\begin{tabular}{ccccccc}
			&                          & N=$2^{10}$                 & N=$2^{11}$                 & N=$2^{12}$                 & N=$2^{13}$                 & N=$2^{14}$                 \\ \hline
			\multicolumn{1}{|c|}{\multirow{3}{*}{1D-DMDT}} & \multicolumn{1}{c|}{CR}  & \multicolumn{1}{c|}{13.50} & \multicolumn{1}{c|}{14.27} & \multicolumn{1}{c|}{14.89} & \multicolumn{1}{c|}{15.30} & \multicolumn{1}{c|}{15.56} \\ \cline{2-7} 
			\cline{2-7} 
			\multicolumn{1}{|c|}{}                         & \multicolumn{1}{c|}{QS}  & \multicolumn{1}{c|}{28.83} & \multicolumn{1}{c|}{30.45} & \multicolumn{1}{c|}{31.96} & \multicolumn{1}{c|}{32.99} & \multicolumn{1}{c|}{33.61} \\ \hline
			\multicolumn{1}{|c|}{\multirow{3}{*}{Ref.\cite{NPaper}}}    & \multicolumn{1}{c|}{CR}  & \multicolumn{1}{c|}{2.90}  & \multicolumn{1}{c|}{4.80}  & \multicolumn{1}{c|}{7.40}  & \multicolumn{1}{c|}{10.50} & \multicolumn{1}{c|}{13.55} \\ \cline{2-7} 
			\cline{2-7} 
			\multicolumn{1}{|c|}{}                         & \multicolumn{1}{c|}{QS}  & \multicolumn{1}{c|}{6.14}  & \multicolumn{1}{c|}{10.25} & \multicolumn{1}{c|}{15.83} & \multicolumn{1}{c|}{22.58} & \multicolumn{1}{c|}{29.19} \\ \hline
		\end{tabular}
	}
	
\end{table}
We simulate the performance of the first subject from the PPG dataset for $N=512$ and thresholding taking values from 5 to 30 with step of 2.  For comparison, we use CDF 9/7 wavelet transform for decomposition because it is considered to be the best for signal compression \cite{6822639}. By using CDF 9/7 wavelet transform, signal is decomposed four times, i.e., level four decomposition of the signal, after that decomposed signal is quantized. Finally, on the quantized signal  Huffman coding is applied. This way of compressing a signal is named WT solution in the paper. The achieved CR and SNR for these thresholds are shown in Fig. \ref{ppg1}. It can be observed that the proposed algorithm performs better than the WT solution for all signal reconstruction quality. For example, WT solution achieves CR=10.1 at SNR= 63.6 dB, whereas the 1D-DMDT compression algorithm achieves CR=15 at SNR=64 db. For this case there is a 48$\%$ gain in CR using 1D-DMDT for SNR around 64 dB than  the WT solution.

\subsection{Accelerometer, Gyroscope, Magnetometer}

\begin{figure}[htbp]
	\centering
	{\includegraphics[width=.8\linewidth]{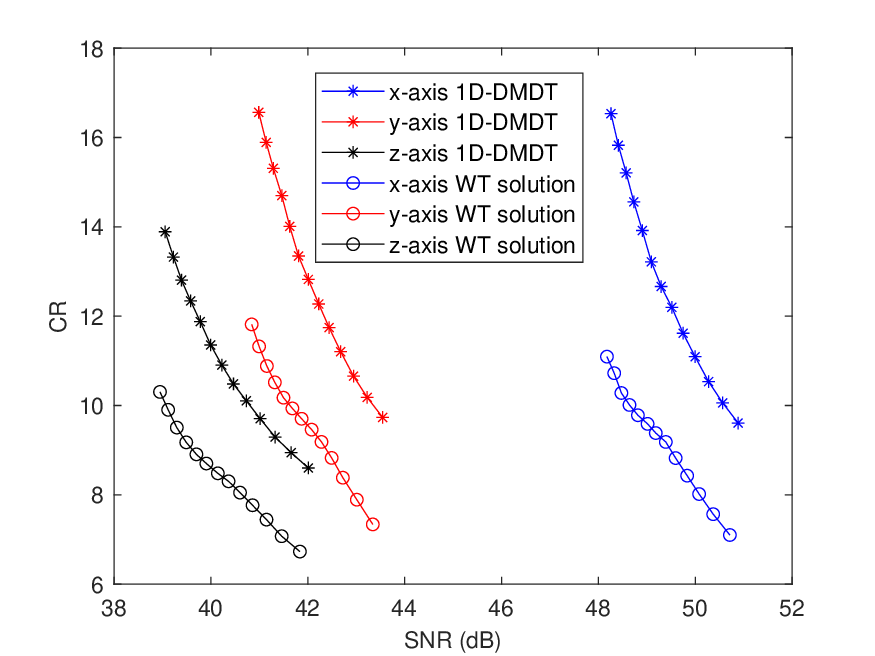}}

	\caption{Performance of 1D-DMDT using accelerometer signals for $N=512$ and various values of thresholds.  For 1D-DMDT first level divisor is 32 and second level divisor is 16. Dataset \cite{BIAGETTI2020105044}}
	\label{acc1}
\end{figure}

\textcolor{black}{In this subsection, we compare 1D-DMDT compression algorithm with WT solution and LFZip for accelerometer, gyroscope, and magnetometer datasets. Simulation results for the WT solution and the 1D-DMDT using accelerometer signal for three axes are shown in Fig. \ref{acc1}. It can be observed that 1D-DMDT compression algorithm performs better than the WT solution for accelerometer data. In particular, compression gain of 1D-DMDT for three axes, i.e., x-axis, y-axis, and z-axis, of accelerometer signal is 37$\%$, 32$\%$, and 28$\%$ as compared to the WT solution. These gains for x-axis, y-axis, and z-axis are observed at SNR 50 dB, 43 dB, and 41 dB, respectively.} 

\begin{figure}[htbp]
	\centering
	{\includegraphics[width=.8\linewidth]{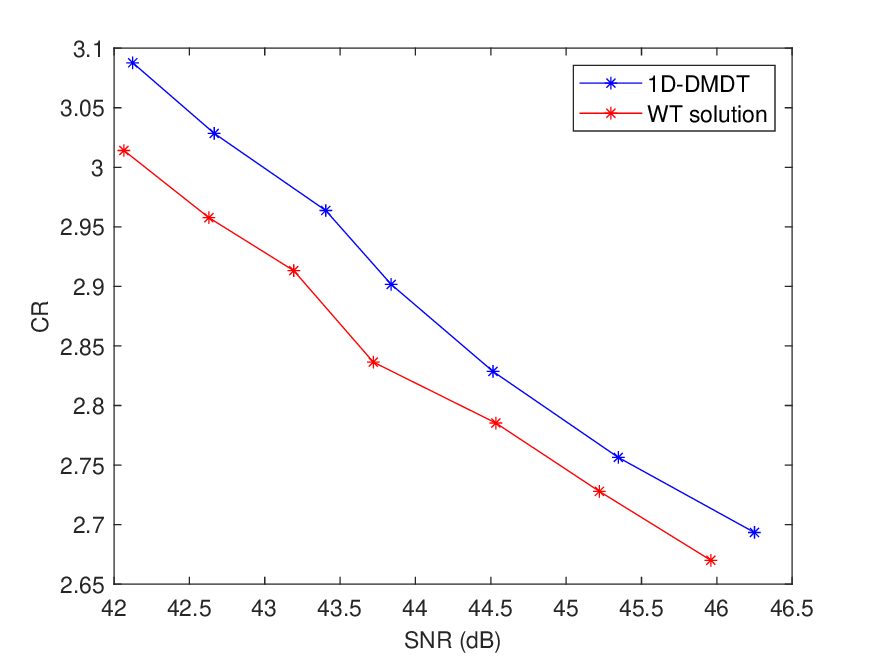}}

	\caption{Performance of 1D-DMDT using magnetometer signals for $N=512$ and various values of thresholds.  For 1D-DMDT first level divisor is 32 and second level divisor is 16. Dataset \cite{CMU1,CMU2}}
	\label{mag1}
\end{figure}

We use magnetometer dataset from \cite{CMU1,CMU2} for comparison and simulations are performed for 1D-DMDT and state-of-the-art WT solution. Simulation results are shown in Fig. \ref{mag1}. It can be observed that 1D-DMDT provides higher CR as compared to WT solution for magnetometer data. The compression gain for magnetometer data is about 5$\%$ at various SNR values.

We also compare 1D-DMDT with LFZip \cite{9105816}. The LFZip is used for compressing floating-point data and it is an error-bounded compression algorithm. In order to compare with LFZip, we use the maximum deviation values defined in LFZip paper \cite{9105816}. We compare CR and distortion allowed in the recovered values for gyroscope dataset and accelerometer dataset \cite{10.1145/2809695.2809718}. All simulations for 1D-DMDT are performed with $d_1=16$ and $d_2=8$, under various values of $\theta$. The maximum deviation values of $10^{-3}$, $10^{-2}$, and $10^{-1}$ are obtained by selecting $\theta=19$,  $180$, and $1500$, correspondingly, for the accelerometer dataset.  For gyroscope dataset $\theta$  is $19$, $220$, and $1500$ for  the maximum deviation values of $10^{-3}$, $10^{-2}$, and $10^{-1}$, respectively. It can be observed from Table \ref{gyroAcce} that the 1D-DMDT achieves higher CR as compared to LFZip for the maximum allowed signal deviations.  The 1D-DMDT compression algorithm provides around 2$\times$ or more improvement in compression ratio for accelerometer and gyroscope signal as compared to the LFZip for deviation $10^{-2}$ or higher.

\begin{table}[]
	\centering
	\caption{Comparison of the 1D-DMDT and LFZip \cite{9105816} for various values of allowed maximum deviation. } \label{gyroAcce}\vspace{0.2cm}
	\begin{tabular}{cc|ccc|}
		\cline{3-5}
		&           & \multicolumn{3}{c|}{ Maximum deviation}                                              \\ \hline
		\multicolumn{1}{|c|}{Dataset}                        & Algorithm & \multicolumn{1}{c|}{$10^{-3}$} & \multicolumn{1}{c|}{$10^{-2}$} & $10^{-1}$ \\ \hline
		\multicolumn{1}{|c|}{\multirow{2}{*}{Accelerometer}} & LFZip     & \multicolumn{1}{c|}{3.55}      & \multicolumn{1}{c|}{5.86}      & 12.71     \\ \cline{2-5} 
		\multicolumn{1}{|c|}{}                               & 1D-DMDT   & \multicolumn{1}{c|}{4.6}       & \multicolumn{1}{c|}{8.64}      & 25.02     \\ \hline
		\multicolumn{1}{|c|}{\multirow{2}{*}{Gyroscope}}     & LFZip     & \multicolumn{1}{c|}{6.05}      & \multicolumn{1}{c|}{12.26}     & 28.77     \\ \cline{2-5} 
		\multicolumn{1}{|c|}{}                               & 1D-DMDT   & \multicolumn{1}{c|}{7.03}      & \multicolumn{1}{c|}{21.48}     & 77.59     \\ \hline
	\end{tabular}
\end{table}

\subsection{Audio}
\begin{figure}[htbp]
	\centering
	\subfloat[]{\includegraphics[width=.8\linewidth]{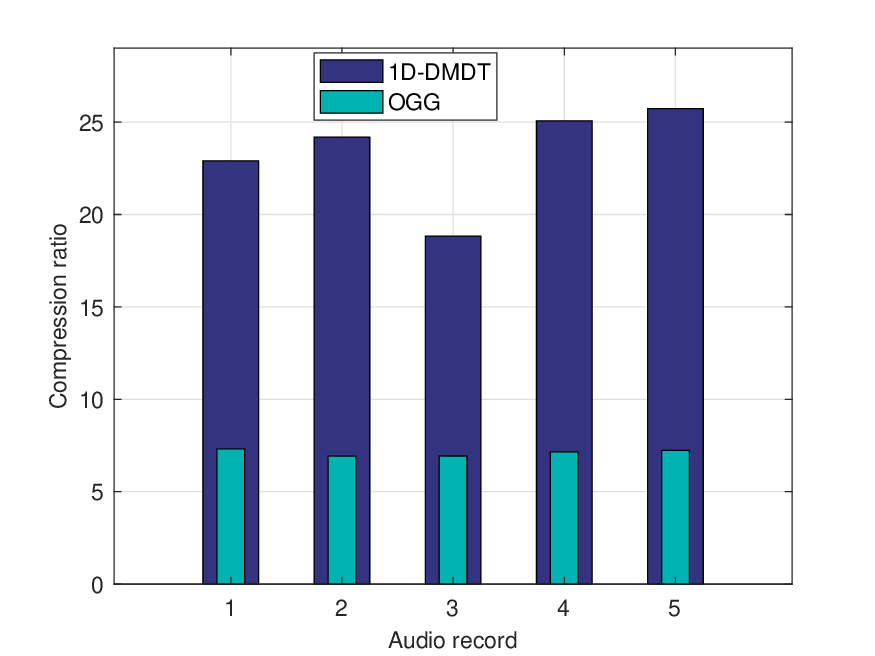}	\label{aud4}}\\
	\subfloat[]{\includegraphics[width=.8\linewidth]{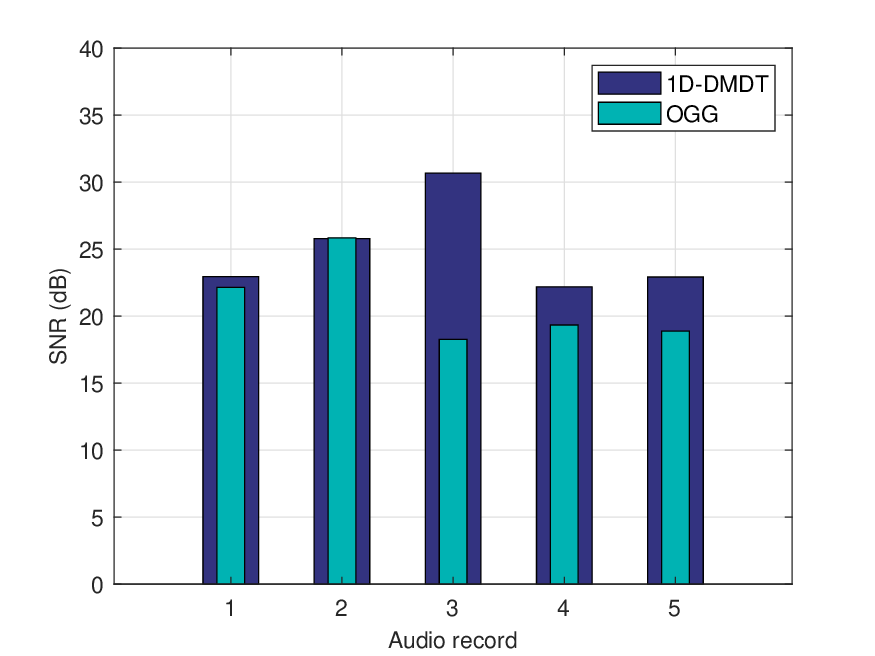} 	\label{aud5}}
	
	\caption{Performance of 1D-DMDT using audio signals for $N=1024$ and various audio records  and  $\theta$=12.  For 1D-DMDT first level divisor is 32 and second level divisor is 16. Dataset \cite{CMU2}. (a) Compression ratio. (b) SNR}
	\label{aud7}
\end{figure}

\begin{figure}[htbp]
	\centering
	{\includegraphics[width=.8\linewidth]{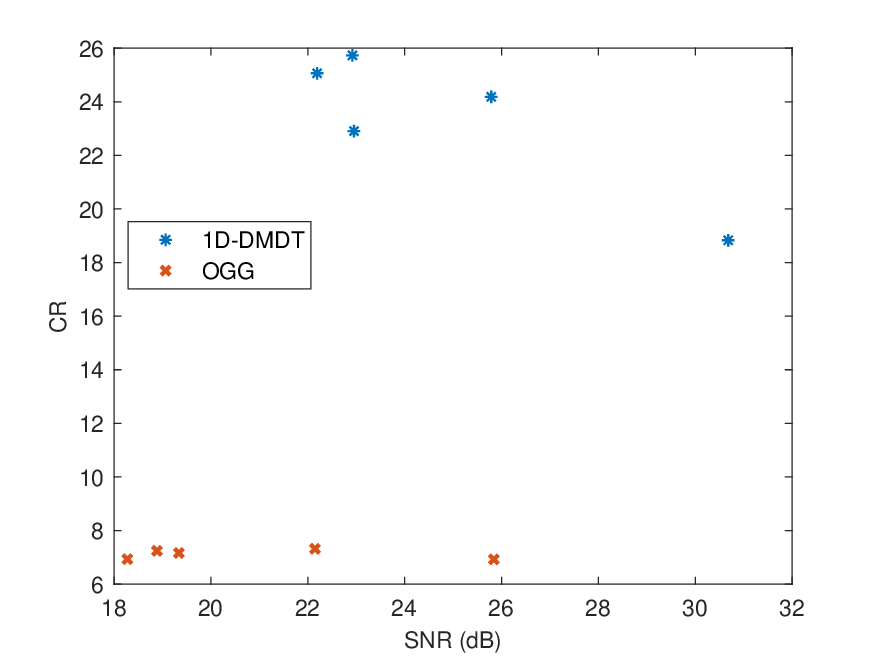}}

	\caption{Performance of 1D-DMDT using audio signals for $N=1024$ and various audio records  and  $\theta$=12.  For 1D-DMDT first level divisor is 32 and second level divisor is 16. Dataset \cite{CMU2}}
	\label{aud6}
\end{figure}

We evaluate the performance of both 1D-DMDT and Vorbis OGG using the audio dataset from \cite{CMU1,CMU2}. Comparison and simulations are performed for both 1D-DMDT and Vorbis OGG. Simulation results for audio are shown in Fig. \ref{aud7} to Fig. \ref{aud6} under different settings. Fig. \ref{aud4} and Fig. \ref{aud5} show that 1D-DMDT achieves 3$\times$ compression gain with better or similar reconstruction signal quality compared to OGG. Fig.\ref{aud6} shows CR and SNR performance for 1D-DMDT and OGG. It can be observed that 1D-DMDT always performs better than OGG for similar SNR values.

%
%

%
%

%
%

\section{Conclusion}
In this paper, we propose a novel discrete multi-level divisor transform. The proposed transform can be applied for any radix and it has multi-resolution property. Furthermore, signal compression algorithm is designed for one-dimensional signals using the DMDT. The proposed algorithm is evaluated for different types of signals, such as, audio, electrocardiogram, accelerometer, magnetometer, photoplethysmography, and gyroscope in comparison with the state-of-the-art compression algorithms. The proposed algorithm achieves $55\%$ or more gain in compression ratio as compared to the  state-of-the-art ECG compression algorithms. The proposed algorithm provides 2-3 times improvement in compression ratio for audio, accelerometer, and gyroscope signals as compared to the state-of-the-art algorithms. Additionally, for accelerometer and photoplethysmography the proposed algorithm achieves 28$\%$ to 48$\%$ compression gain than wavelet based compression algorithm at similar reconstruction quality.

\bibliographystyle{ieeetr}
\bibliography{Bibliography.bib}

\end{document}